\documentclass[12pt]{article}
\usepackage{amsmath}
\usepackage{feynmf}
\newcommand{\Tr}{\ensuremath{\mathop{\mathrm{Tr}}}}
\newcommand{\tr}{\ensuremath{\mathop{\mathrm{tr}}}}

\newcommand{\del}{\partial}

\begin{document}


\title{ Wavy Wilson Line and AdS/CFT}

\author{Gordon W. Semenoff and Donovan Young\\
Department of Physics and Astronomy, University of British
Columbia,\\ Vancouver, British Columbia, Canada V6T
1Z1\\dyoung@physics.ubc.ca~~semenoff@nbi.dk }


\maketitle


\begin{abstract}
Wilson loops which are small deviations from straight, infinite
lines, called wavy lines, are considered in the context of the
AdS/CFT correspondence. A single wavy line and the connected
correlation function of a straight and wavy line are considered.
It is argued that, to leading order in ``waviness'', the functional
form of the loop is universal and the coefficient, which is a
function of the 't Hooft coupling, is found in weak coupling
perturbation theory and the strong coupling limit using the
AdS/CFT correspondence. Supersymmetric arguments are used to
simplify the computation and to show that the straight line obeys
the Migdal-Makeenko loop equation.

\end{abstract}



\newpage\setcounter{page}{1}
\begin{fmffile}{big}

\section{Introduction and Summary}

The AdS/CFT correspondence has provided a fascinating array of
relationships between gauge field theories, string theory and
supergravity\cite{Maldacena:1997re}-\cite{Aharony:1999ti}. One of
the natural objects of gauge theory which couples directly to
strings is the Wilson loop. The study of Wilson loops has provided
an interesting approach to extracting information from the AdS/CFT
correspondence \cite{Maldacena:1998im}-\cite{Kotikov:2004er}.

In ${\cal N}=4$ supersymmetric Yang-Mills theory\footnote{The
action and other conventions are summarized in the Appendix.}, the
Wilson loop of most interest contains both the gauge field and a
scalar field in the exponent and has the form in Euclidean
space\cite{Maldacena:1998im}
\begin{equation}\label{wloop} {\rm Tr}\left({\cal P}~ e^{\oint
ds\left(i A_\mu(x)\dot x^\mu(s)+\Phi(x)\cdot\theta|\dot
x(s)|\right)} \right)
\end{equation}
This loop measures the holonomy of the wave-function of a heavy
W-boson which occurs when the gauge symmetry of super Yang-Mills
theory is realized in a Higgs phase with the unit vector
$\theta^I$ related to the condensate, $\left<\Phi^I(x)\right>\sim
\theta^I$.

Computations of the expectation value of this Wilson loop  have
proven to be tractable in certain special geometries. For example,
the infinite straight line, or any array of infinite parallel
straight lines form a BPS object and it is expected that all
radiative corrections cancel, so that the Wilson loop
corresponding to them has expectation value exactly equal to one.

The expectation value of the circular loop, which is also a BPS
object closely related to the straight line, is conjectured to be
known exactly\cite{Erickson:2000af}. In that case ladder diagrams
can be summed explicitly. The sum can be extrapolated to strong
coupling and compared with the predictions of the AdS/CFT
correspondence where it agrees beautifully. It is conjectured that
all corrections to ladder diagrams cancel.  This has been
demonstrated to leading and next-to-leading orders
\cite{Erickson:2000af,Plefka:2001bu,Arutyunov:2001hs} and there
are other arguments to support it\cite{Drukker:2000rr}. Similar
observations have been made for the correlators of chiral primary
operators with the circular Wilson loop\cite{Semenoff:2001xp}.

Polyakov and Rychkov
\cite{Polyakov:2000ti,Kazakov,Polyakov:2000jg} have discussed
Wilson loops which were small deviations from straight lines,
their so-called ``wavy lines''. There, they observed some
interesting structures which gave some hope that the area Ans\"atz
at strong coupling actually satisfied the loop equations of the
gauge theory.

In this Letter, we shall present some preliminary results of our
investigation of wavy lines.   We begin by reviewing some
preliminaries.

\subsection{Preliminaries}

We will be entirely concerned with Wilson loops in
four-dimensional Euclidean space. There is some closely related
and very interesting work on Minkowski space
loops\cite{Mikhailov:2003er}. Apparent differences between those
and the present work are attributable to the richer array of
boundary conditions which can be imposed in Minkowski space.

A wavy line deviates by a small amount from an infinite straight
line. We shall describe it using the Monge gauge parameterization
\begin{equation}x^\mu(s)=(s, \xi(s))  ~~~s\in(-\infty,
+\infty)\label{par}\end{equation} The three-dimensional vector
$\xi(s)$ is a smooth function of the curve parameter $s$ with
small magnitude.

The expectation value of the Wilson loop is a functional of the
geometry of the loop.  We will consider the leading order of this
functional, which is quadratic in $\xi(s)$. This leading order is
restricted by the spacetime symmetries of ${\cal N}=4$
supersymmetric Yang-Mills theory.  Rotation and translation
invariance dictate that it has the form
\begin{equation} \int ds\int ds'
\dot\xi^i(s)K(s-s')\dot\xi^i(s')\label{gaugeterm}\end{equation}
Scale invariance indicates that the kernel $K(s-s')$ has dimension
1/distance$^2$ and is therefore of the form $$K(s-s')\sim
\frac{1}{(s-s')^2} $$ However the integration then diverges
linearly.  We do not expect that such divergences appear in the
supersymmetric Wilson loop that we are considering. Therefore, the
kernel must be a distribution. There are two distributions with
the correct dimensions,
$$
\frac{d}{ds}\delta(s-s') ~~,~~\frac{d}{ds}\frac{\bf P}{s-s'}
$$
The second of these, the derivative of the principal value
distribution, is an even function, so we must chose it. By adding
terms which integrate to zero, we can then write the functional in
a more manifestly finite form,
\begin{equation}
 \int ds \int ds' \left\{
2\dot\xi(s)\cdot\dot\xi(s')-2\dot\xi(s')^2\right\}
 \frac{d}{ds}\frac{\bf
P}{(s-s') }~=~ \int ds \int ds' \frac{
\left(\dot\xi(s)-\dot\xi(s')\right)^2}{(s-s')^2 }
\label{fullterm}\end{equation}
 We shall see that this is precisely
the form that we obtain for the wavy line, both in perturbation
theory and in the strong coupling limit using AdS/CFT.  In the
leading order in perturbation theory, the part in
(\ref{gaugeterm}) comes from the gauge interactions, whereas the
extra terms needed to make (\ref{fullterm}) come from the scalar
fields.

In addition, we shall consider the connected correlation function
of a wavy line with a straight line.  In that case, the correlator
depends on the distance, $L$, between the two lines. We will
compute the leading order, which varies as $\frac{1}{L^2}$. Then,
the same reasoning and counting of dimensions tells us that the
correlator must be of the form
\begin{equation}
\label{straightwavy} \frac{1}{L^2}\int ds \int
ds'\dot\xi(s)\cdot\dot\xi(s')\ln[\Lambda^2(s-s')^2]
\end{equation}
Here, $\Lambda$ is a constant with the dimension of inverse
length.

\subsection{Results}

Our results can be summarized as follows:
\begin{itemize}
\item{}Unlike the circle and other loops that have been computed
in the past, where ladder diagrams were the most important, the
wavy line gets all of its corrections from internal loops. Any
ladder diagrams either cancel or vanish identically (beyond the
trivial leading order for the single wavy line).

 \item{}We shall find that the wavy line and the connected
 correlation function of a wavy line with a
straight line indeed have the universal forms, (\ref{fullterm})
and (\ref{straightwavy}), respectively. We show this to leading
order in weak coupling perturbation theory and we confirm it at
strong coupling using the AdS/CFT correspondence.  For the single
wavy line, we also confirm that it is so to next-to-leading order
at weak coupling.

\item{}In the universal form (\ref{fullterm}), to leading orders
in perturbation theory, the role of the scalar field in the Wilson
loop is minimal. It serves to regulate divergences and define the
distribution in the kernel.  This is consistent with the results
of Polyakov and Rychkov\cite{Polyakov:2000ti,Polyakov:2000jg} who
applied similar ideas to non-supersymmetric loops.

\item{}The coefficients of the universal functionals are
nontrivial functions of the coupling constant which we expect
obtain contributions from all orders in perturbation theory.

\item{}One way that a scale dependence could creep into the Wilson
loop is if the power law in (\ref{fullterm}) is corrected by
logarithms in higher orders of perturbation theory. In the case of
the wavy line, we shall confirm to the next to leading order that
it is not corrected by logarithms. We also use AdS/CFT to compute
the strong coupling limit and find the same functional form,
suggesting that logarithms do not appear at any order.

\item{}The above statement is even more interesting in the case of
(\ref{straightwavy}) where, in our explicit computations, a
logarithm of the cutoff $\Lambda$ in fact appears.   The integral,
however, is insensitive to the appearance of this logarithm.  It
can be removed by adding a term which is a total derivative.  If
higher orders in logarithms appeared there, it is hard to see how
the cutoff dependence could be removed in this way.  We confirm
using AdS/CFT that, at strong coupling, there is indeed only this
single logarithm.  In that case, the ultraviolet cutoff,
$\Lambda$, is replaced by the inverse of an infrared cutoff, a
symptom of the interchange of ultraviolet and infrared behaviors
which occurs in the AdS/CFT correspondence in general.

\item{}We use supersymmetry to simplify the computation of
correlation functions and put them in a form where further
computations can be done more readily.

\item{}We use supersymmetry to find that the infinite straight
line obeys the Migdal-Makeenko\cite{Makeenko:pb,Makeenko:vm}loop
equation of gauge theory.  This is beautifully consistent with the
results of and Drukker, Gross and Ooguri\cite{Drukker:2000rr} and
Polyakov and Rychkov\cite{Polyakov:2000ti,Polyakov:2000jg}  who
showed that the strong coupling Ans\"atz obeys the loop equation.
For a wavy line, this can actually be deduced directly from the
fact that it has the functional form in (\ref{fullterm}) where the
kernel does not contain the delta function singularity which would
be identified with the loop operator in the quadratic variation of
the loop.

\item{}A local limit of the waviness can be taken so that one
could in principle use the operator product expansion to compute
the correlation functions of gauge invariant operators with the
straight line Wilson loop.  We hope to report results in the near
future.
\end{itemize}

\section{Weak Coupling}

\subsection{Single Line}

We have calculated the expectation value of a single wavy line
to second order in the 't Hooft coupling. The calculation involves
evaluating various Feynman diagrams. In the following the horizontal
line denotes the Wilson line, the wiggly line denotes the gauge field $A_\mu$,
while the solid line denotes the scalar field $\Phi_J$.

\[
\parbox{20mm}{
\begin{fmfgraph}(20,20)
\fmfleft{i}
\fmfright{o}
\fmf{plain}{i,v1}
\fmf{wiggly,left,tension=.2}{v1,v2}
\fmf{plain,tension=.2}{v1,v2}
\fmf{plain}{v2,o}
\end{fmfgraph}}\,
\hspace{1cm}+\hspace{1cm}
\parbox{20mm}{
\begin{fmfgraph}(20,20)
\fmfleft{i}
\fmfright{o}
\fmf{plain}{i,v1}
\fmf{plain,left,tension=.2}{v1,v2}
\fmf{plain,tension=.2}{v1,v2}
\fmf{plain}{v2,o}
\end{fmfgraph}}\,
\hspace{0.6cm}=~~~\frac{g^2N}{16\pi^2}I
\]
\[
\parbox{20mm}{
\begin{fmfgraph}(20,20)
\fmfleft{i}
\fmfright{o}
\fmf{plain}{i,v1}
\fmf{plain}{v1,v2}
\fmf{plain}{v2,v3}
\fmf{plain}{v3,v4}
\fmf{plain}{v4,v5}
\fmf{plain}{v5,o}
\fmffreeze
\fmf{wiggly,left, tension=2}{v2,v4}
\fmf{wiggly,left, tension=2}{v1,v5}
\end{fmfgraph}}\,
\hspace{0.3cm}+\hspace{0.3cm}
\parbox{20mm}{
\begin{fmfgraph}(20,20)
\fmfleft{i}
\fmfright{o}
\fmf{plain}{i,v1}
\fmf{plain}{v1,v2}
\fmf{plain}{v2,v3}
\fmf{plain}{v3,v4}
\fmf{plain}{v4,v5}
\fmf{plain}{v5,o}
\fmffreeze
\fmf{plain,left, tension=2}{v2,v4}
\fmf{wiggly,left, tension=2}{v1,v5}
\end{fmfgraph}}\,
\hspace{0.3cm}+\hspace{0.3cm}
\parbox{20mm}{
\begin{fmfgraph}(20,20)
\fmfleft{i}
\fmfright{o}
\fmf{plain}{i,v1}
\fmf{plain}{v1,v2}
\fmf{plain}{v2,v3}
\fmf{plain}{v3,v4}
\fmf{plain}{v4,v5}
\fmf{plain}{v5,o}
\fmffreeze
\fmf{wiggly,left, tension=2}{v2,v4}
\fmf{plain,left, tension=2}{v1,v5}
\end{fmfgraph}}\,
\hspace{0.3cm}+\hspace{0.3cm}
\parbox{20mm}{
\begin{fmfgraph}(20,20)
\fmfleft{i}
\fmfright{o}
\fmf{plain}{i,v1}
\fmf{plain}{v1,v2}
\fmf{plain}{v2,v3}
\fmf{plain}{v3,v4}
\fmf{plain}{v4,v5}
\fmf{plain}{v5,o}
\fmffreeze
\fmf{plain,left, tension=2}{v2,v4}
\fmf{plain,left, tension=2}{v1,v5}
\end{fmfgraph}}\,\hspace{0.3cm}=0\]
\[\parbox{20mm}{
\begin{fmfgraph}(19.5,5.625)
\fmfstraight
\fmfbottom{i1,i2,i3,i4,i5,i6,i7}
\fmftop{o1,o2,o3,o4,o5,o6,o7}
\fmf{plain}{i1,i2,i3,i4,i5,i6,i7}
\fmf{wiggly,left=0.4}{i2,o3}
\fmf{wiggly,left=0.4}{o5,i6}
\fmfv{d.sh=circle,d.filled=empty,d.size=80}{o4}
\end{fmfgraph}}\,
\hspace{0.3cm}+\hspace{0.3cm}
\parbox{20mm}{
\begin{fmfgraph}(19.5,5.625)
\fmfstraight
\fmfbottom{i1,i2,i3,i4,i5,i6,i7}
\fmftop{o1,o2,o3,o4,o5,o6,o7}
\fmf{plain}{i1,i2,i3,i4,i5,i6,i7}
\fmf{plain,left=0.4}{i2,o3}
\fmf{plain,left=0.4}{o5,i6}
\fmfv{d.sh=circle,d.filled=empty,d.size=80}{o4}
\end{fmfgraph}}\,
\hspace{0.3cm}+\hspace{0.3cm}
\parbox{20mm}{
\begin{fmfgraph}(19.5,5.625)
\fmfstraight
\fmfbottom{i1,i2,i3,i4,i5,i6,i7}
\fmftop{o1,o2,o3,o4,o5,o6,o7}
\fmf{plain}{i1,i2,i3,i4,i5,i6,i7}
\fmf{plain,left=0.4}{i2,o4,i6}
\fmf{wiggly}{i4,o4}
\fmfdot{o4}
\end{fmfgraph}}\,
\hspace{0.3cm}+\hspace{0.3cm}
\parbox{20mm}{
\begin{fmfgraph}(19.5,5.625)
\fmfstraight
\fmfbottom{i1,i2,i3,i4,i5,i6,i7}
\fmftop{o1,o2,o3,o4,o5,o6,o7}
\fmf{plain}{i1,i2,i3,i4,i5,i6,i7}
\fmf{wiggly,left=0.4}{i2,o4,i6}
\fmf{wiggly}{i4,o4}
\fmfdot{o4}
\end{fmfgraph}}\,\hspace{0.6cm} =~~~
- \frac{g^4 N^2}{2^7 \pi^2}\frac{1}{3} I
\]\\

\noindent where,

\begin{equation}
I = \oint ds_1\,ds_2
\frac{\left[ \dot\xi(s_1) - \dot\xi(s_2) \right]^2}{2 \left(s_1-s_2\right)^2}
\label{eye}
\end{equation}

\noindent The first two diagrams in the last line represent the
one-loop corrected exchange of a single particle. These diagrams
are divergent, but a divergent piece from the following diagrams
(those with an internal vertex) cancel these divergences exactly,
leaving a finite result. The diagrams in the second line are zero
at second order in waviness individually. Summarizing, we find

\begin{equation}
\langle W(C)\rangle = 1 + \left[\frac{g^2 N}{2^4 \pi^2} -
\frac{g^4 N^2}{3\cdot 2^7 \pi^2}+\ldots \right] \oint ds_1\,ds_2
\frac{\left[ \dot\xi(s_1) - \dot\xi(s_2) \right]^2}{2
\left(s_1-s_2\right)^2}+\ldots. \label{singlefinal}
\end{equation}

\subsection{Line-Line Correlator}

A Wilson loop which consists of an array of infinite parallel
straight lines is a BPS object.  As an operator it commutes with
half of the supercharges and one might expect that, like the
single straight line, it is protected from quantum corrections.
Explicit computations to a few orders in perturbation theory
indeed show that lower order corrections cancel and one might
conjecture that they do to all orders.  This is consistent with
what is found in the strong coupling limit using the AdS/CFT
correspondence.  It has the physical interpretation that an array
of static heavy quarks do not interact.

An interesting variant of this configuration is a combination of
an infinite straight line and a wavy deformation of another
parallel straight line. In this case, the connected correlation
function measures the interaction energy of two quarks which is
induced by the slight motion of one of them.

We can compute the correlation function of an infinite straight
line and a wavy line in perturbation theory. The leading
contribution turns out to be at cubic order in the 't Hooft
coupling and is given by the following diagram:

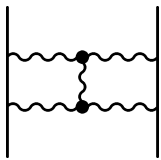
\begin{center}
\parbox{20mm}{
\begin{fmfgraph*}(20,20)
\fmfstraight
\fmfpen{thin} \fmfleft{i1,i2,i3,i4} \fmfright{o1,o2,o3,o4}
\fmf{plain}{i1,i4}
\fmf{plain}{o1,o4}
\fmffreeze
\fmf{wiggly}{i2,v1,o2}
\fmffreeze
\fmf{wiggly}{i3,v2,o3}
\fmffreeze
\fmf{wiggly}{v1,v2}
\fmfdot{v1,v2}
\end{fmfgraph*}}
\end{center}

\noindent where the two Wilson loops are represented as vertical
lines.  The other ``H-diagrams'' obtained by all possible
substitutions of scalar for gauge fields also contribute.

In the limit of large separation $L$, we find:

\begin{equation}
\left< W_1 W_2 \right>_{\rm connected} =  - \frac{1}{N^2} \,
\frac{1}{L^2} \, \left[\frac{g^6 N^3}{2^{11} \pi^4}+\ldots\right]
\, \oint \, ds \, ds' \, {\dot \xi}(s) \cdot {\dot \xi}(s') \, \ln
( \Lambda^2 (s-s')^2 )+\ldots \label{pert_final_L}
\end{equation}

The parameter $\Lambda$ in the logarithm is an ultraviolet cutoff.
Note that the result is finite.  The cutoff disappears when we
integrate by parts in $s$ or $s'$.

\section{Strong Coupling}

The AdS/CFT correspondence hypothesizes that the strong coupling
limit of the Wilson loop is found by computing the regularized
minimal area of a surface in $AdS_5\times S^5$ whose boundary is
the curve of the Wilson loop, embedded in the boundary of the
space\cite{Maldacena:1998im}.

 We use the metric of $AdS_5\times S^5$,
 $$
 ds^2=R^2 \frac{ dx^\mu dx^\mu + dy^I dy^I}{y^I y^I}
 $$
 where, according to the AdS/CFT correspondence, the radius of curvature is
 $R=(g^2N)^{1/4}\sqrt{\alpha'}$.
 We shall consider surfaces which are located on a single point in $S^5$, given
 by $y^I=y\theta^I$.  The boundary of the space is located at $y\to 0$.

\subsection{Single Line}

The straight line is the boundary of a surface which is orthogonal
to the boundary of AdS, which we parameterize using the
coordinates $(s,t)$.  The parametric embedding of the surface is
\begin{equation}
(x^\mu,y^I)=(s,0,0,0,t\theta^I) \label{straight}\end{equation}
This surface is itself $AdS_2$, with metric
$$
ds^2= R^2 \frac{ ds^2+ dt^2}{t^2}
$$
It has infinite area,
$$
{\cal A}_0= \int_{-\infty}^\infty ds\int_0^\infty dt
\frac{R^2}{t^2}
$$
which must be defined by regularization and the infinity must be
subtraced to obtain the final result. The regularization is
normally carried out by cutting off the integrals
$$
{\cal A}_0= \int_{-L/2}^{L/2} ds\int_\epsilon^\infty dt
\frac{R^2}{t^2}= R^2 L/\epsilon
$$
The subtraction of the infinite part is implemented by operating
$\left(1+\epsilon\frac{d}{d\epsilon}\right)$ which has an
interesting interpretation as a Legendre
transform\cite{Drukker:1999zq,Semenoff:2002kk}.  In the case of an
infinite straight line this subtracts everything
$$
{\cal A}_{0{\rm
Reg}}=\left(1+\epsilon\frac{d}{d\epsilon}\right)R^2\frac{L}{\epsilon}=0
$$
so the result is
$$
\left< W({\rm
straight~line})\right>=e^{-\frac{R^2}{2\pi\alpha'}{\cal A}_{0{\rm
Reg}}}=1
$$

To describe a wavy line, we must find a minimal surface whose
boundary is the wavy line.  We describe the surface using the
embedding coordinates
\begin{equation}
(x^\mu, y^I) = (s,\, \Delta_j(t,s), \,t\theta^I) \quad
\text{where} \quad \Delta_j(0,s) = \xi_j(s) \label{embed}
\end{equation}
The three components $\Delta_j$ are the small deviation from
$AdS_2$ induced by the waviness of the line.

 The regularized area to second order in $\Delta$ is given by

\begin{equation}
{\cal A}_{\rm Reg} = -\int \frac{dt \, ds}{t^2} \left\{
\frac{1}{2} (\del_t \Delta)^2 + \frac{1}{2} (\del_s\Delta)^2
\right\} \label{S2}
\end{equation}

\noindent The variation of this functional gives an equation of
motion for $\Delta$, the solution of which, with the boundary
condition in (\ref{embed}) is
\begin{equation}
\Delta(t, s) = \int \frac{ds'\xi(s')}{\pi} \frac{2t^3}{((s-s')^2 +
t^2)^2}
\end{equation}
When plugged back into (\ref{S2}) we find the area

\begin{equation}
{\cal A}_{\rm Reg}[\xi] = -\frac{1}{2\pi} \int ds' ds
\frac{\left[{\dot \xi}(s) - {\dot \xi}(s')\right]^2}{2(s-s')^2}
\label{string1}
\end{equation}
and  the strong coupling limit of the wavy line Wilson loop is
$$
\left<W(-)\right> = \exp\left( \frac{\sqrt{g^2N}}{4\pi^2}\int ds'
ds \frac{\left[{\dot \xi}(s) - {\dot
\xi}(s')\right]^2}{2(s-s')^2}+\ldots \right)
$$
Note that, as was expected, this procedure gives the same
functional of the deviation from the straight line as we found at
weak coupling.  This supports the idea that the power law behavior
is not corrected by logarithms at intermediate orders in
perturbation theory.  The coefficient seems to be a non-trivial
function of the coupling, interpolating between a linear function
at weak coupling and the square root at strong coupling.

\subsection{Line-Line Correlator}

If we consider an array of parallel straight lines, they are the
boundaries of a set of sheets in $AdS_5\times S^5$ similar to
(\ref{straight}).  If two lines are anti-parallel, they can be
joined by a single sheet which dominates their connected
correlation function. Since, in the case of interest to us, the
lines are parallel, rather than anti-parallel, there is no single
sheet whose boundary is more than one of the lines. This means
that in the leading order, the lines do not interact, i.e. their
connected correlation function vanishes.  This is consistent with
the weak coupling expansion where we saw that the connected
correlation function indeed has a coefficient proportional to
$1/N^2$ (times a function of the 't Hooft coupling $g^2N$) which
indicates that it arises from higher genus Feynman diagrams. In
the strong coupling limit, we expect this to translate to higher
genus surfaces.

Still, we expect that, for exactly parallel lines, because of
their BPS nature, higher genus contributions cancel exactly.  We
emphasize that we do not know an explicit proof of this statement.
It can be checked to leading orders in weak coupling and it
appears to occur there.  However, at strong coupling, we are not
even able to check it to the leading order, but we shall assume
that it is the case.

It is possible to take into account higher genus surfaces to the
straight line-wavy line correlation function if we consider the
limit where the lines are far apart compared to the distance scale
of the waviness. In this case, at higher genus, the lines are
boundaries of two infinite sheets which are connected by  thin
tubes, which are formed by the exchange of the light particles in
the spectrum of supergravity linearized about the $AdS_5\times
S^5$ background.

This idea was used  in \cite{Berenstein:1998ij} to compute the
correlator of widely separated circular Wilson loops.     The
Euler character $\chi=-2$ worldsheet then has the area

\begin{equation}
{\cal A}_{\chi=-2}= \frac{g^2N}{4\pi^2}\int_{\Omega_1}
\int_{\Omega_2} {\cal V}_1 \, P \, {\cal V}_2 \label{form}
\end{equation}

\noindent where $\Omega_i$ refers to the worldsheet domains,
${\cal V}_i$ refers to the vertex operator on that worldsheet, and
$P$ denotes the propagator for the supergravity field which
travels between the worldsheets.  The factor in front comes from
two powers of the coefficient of the worldsheet action,
$(R^2/2\pi\sqrt{\alpha'})^2=g^2N/4\pi^2$.

The lightest mode comes from the perturbation of the $AdS_5$
metric, which is expressed in terms of a Kaluza-Klein scalar (in
the following equation greek indices refer to the five coordinates
on $AdS_5$)

\begin{equation}
{\tilde g}_{\mu\nu} = g_{\mu\nu} -\frac{10k}{3} g_{\mu\nu} \, s^k \, Y^k
+ \frac{4}{k+1} D_{(\mu} D_{\nu)} \, s^k \, Y^k +
\frac{32k}{15} g_{\mu\nu} \, s^k \, Y^k,
\label{delta_g}
\end{equation}

\noindent where $Y^k$ is the spherical harmonic on $S^5$, and
$s^k$ is the Kaluza-Klein scalar field described in section 4.2 of
\cite{Berenstein:1998ij}. The lightest mode of $s^k$ is for $k=2$.
The resulting perturbation in the Nambu action on $AdS_5$ yields
the vertex operator. At zeroth order in waviness the vertex
operator is

\begin{equation}
{\cal V}_{\Delta = 0} = -\frac{12}{5} \frac{1}{t^2} -\frac{4}{15} {\vec \nabla}^2
+\frac{4}{5} \frac{1}{t} \del_t + \frac{2}{5} (\del_t^2 + \del_s^2)
\end{equation}

\noindent where $s$ and $t$ are the embedding coordinates as in
(\ref{embed}), i.e. the worldsheet coordinates. The derivatives
will act on the propagator. In the limit of large separation the
propagator is

\begin{equation}
P = \left. \frac{9}{8N^2} \frac{t^2 t'^2}{\left[(s-s')^2 + (t-t')^2 +
({\vec x} - {\vec x}')^2\right]^2} \right|_{{\vec x} = {\vec x}'+ {\vec L}}
\end{equation}

\noindent where the separation is given by $\vec L$. As for the vertex operator
of the wavy line, i.e. the vertex operator terms which are second order in waviness,
we keep only those terms which contain derivatives in $t'$, other terms producing
results subleading in the separation. We find

\begin{equation}
\begin{split}
{\cal V}(s',t') &= \left( \frac{1}{2} {\dot \Delta}^2 +
\frac{1}{2} \Delta'^2 \right) \frac{1}{t'^2} \left\{ -\frac{12}{5}
- \frac{4}{15} \left[t'^2 \del_{t'}^2
-3t'\del_{t'} \right] \right\}\\
&+ \frac{1}{3} \left( -{\dot \Delta}^2 \del_{t'}^2 + \Delta'^2 \del_{t'}^2 \right)
- \frac{4}{3} \frac{1}{t'} {\dot \Delta}^2 \del_{t'}
\end{split}
\end{equation}

\noindent where $\Delta'(t',s') = \del_{s'}\Delta(t',s')$ and
${\dot \Delta}(t',s') = \del_{t'}\Delta(t',s')$, and
where derivatives act on the numerator of the propagator only, else
leading to subleading terms. Plugging everything into (\ref{form}), we find
to leading order in the separation

\begin{equation}
{\cal A}_{\chi=-2}= \frac{1}{L^2}\frac{1}{N^2}\frac{g^2N}{8\pi^2}
\frac{3}{20} \int ds \, ds' \, {\dot \xi(s)} \cdot {\dot \xi(s')}
\, \ln ( \mu^2 \, (s-s')^2 ) \label{strong_final_L}
\end{equation}

This has the same dependence on $\xi$ as the weak coupling limit,
supporting our expectation that the logarithm is not modified by
loop corrections and it remains the same universal form at all
orders.  The coefficient is indeed of order $1/N^2$ and the
coupling constant appears to be nontrivial.

\section{Supersymmetry, Simplifications, and the Loop Equation}
\label{susysection}

Consider the expansion of the single wavy line to two orders in
the function $\xi$,
\begin{equation}
\begin{split}
\delta^2 W(C) = &\frac{1}{N} \Tr \int ds \int dt ~ {\cal P}
e^{\int_s^\infty ds_1 E(s_1) } {\cal O}_1(s) {\cal P} e^{\int_t^s
ds_2 E(s_2) } {\cal O}_1(t)
{\cal P} e^{\int_{- \infty}^t ds_3 E(s_3) }\\
+ &\frac{1}{N} \Tr \int ds {\cal P} e^{ \int_s^\infty ds_1 E(s_1)
} {\cal O}_2(s) {\cal P} e^{ \int_{ - \infty }^s ds_2 E(s_2) },
\label{2opins}
\end{split}
\end{equation}

\noindent where,

\begin{equation}
\label{exponent} E(s) = i A_0(x(s)) + \Phi(x(s)) \cdot \theta
\end{equation}

\begin{equation}
{\cal O}_1(s) = \xi_j(s) \left[\,i F_{j 0}(x(s)) + D_j \Phi(x(s))
\cdot \theta \,\right] \label{O1}
\end{equation}

\begin{equation}
{\cal O}_2(s) = \xi_k(s) \xi_j(s) \left[\, i D_j F_{k 0}(x(s)) +
D_j D_k \Phi(x(s)) \cdot \theta \,\right]. \label{O2}
\end{equation}

\noindent In the following we will show that the second term in
(\ref{2opins}) is the result of acting the loop space Laplace
operator on the Wilson loop and  vanishes identically.

We shall also find that the wavy line operators (\ref{O1},
\ref{O2}) can be described, via supersymmetry, in terms of
fermionic operators. This provides a simplification of the Feynman
diagrams involved in the various computations, as well as a proof
that (\ref{O2}) vanishes to all orders in perturbation theory,
which means that the straight line Wilson loop obeys the loop
equation.

We use the ten dimensional supersymmetry transformations, as
$D=4,\,{\cal N}=4$ super-Yang-Mills theory is just a dimensional
reduction of $D=10,\,{\cal N}=1$ super-Yang-Mills theory. The
supersymmetry transformations are

\begin{equation}
\delta A^\mu = \frac{i}{2} {\bar \epsilon} \gamma^\mu \psi
\quad \,
\delta \Phi^m = \frac{i}{2} {\bar \epsilon} \Gamma^m \psi
\quad \,
\delta \psi = -\frac{1}{4} \Gamma^{MN} F_{MN} \epsilon
\quad \,
\Gamma^{MN} = \frac{1}{2} [ \Gamma^M, \Gamma^N ]
\end{equation}

\noindent where $M = (0,i,m)$ so that $i = 1,2,3$, $m = 4,...,9$, and
$\mu = 0,...,3$. The 10-D gamma matrices are $\Gamma^M = (\gamma^\mu, \Gamma^m)$
and $\psi$ is a 10-D Majorana-Weyl fermion. The generalized field strength
$F^{MN}$ is understood as being built from the 10-D gauge field
$A^M = (A^\mu, \Phi^m)$.

We identify a projected supersymmetry transformation which
commutes with the straight line Wilson loop, that is with the
exponent (\ref{exponent})

\begin{equation}
{\bar \epsilon} = {\bar \eta} \, (i \gamma^0 + \Gamma \cdot \theta).
\end{equation}

\noindent Let the supercharge responsible for this subset of transformations
be called $Q_p$, while the full supercharge we will call $Q$. We find that

\begin{equation}
{\cal O}_1 = \frac{i}{4}
\tr \left( \left\{ Q_p, \left[ {\bar Q}, \xi_i A_i \right] \right\} \right)
\qquad
{\cal O}_2 = \frac{i}{4}
\tr \left( \left\{ Q_p, \left[ {\bar Q}, \xi_i \xi_j D_{(i}A_{j)} \right] \right\} \right)
\end{equation}

\noindent where the trace is over Dirac indices. This allows us to write the
following identity

\begin{equation}
\begin{split}
0 &= \frac{i}{4} \tr \left< \left\{Q_p,\frac{1}{N} \Tr \int ds\,dt ~
{\cal P} e^{\int E} \,
{\cal O}_1(s) \, {\cal P} e^{\int E} \, [{\bar Q},A_i](t) \,
{\cal P} e^{\int E}  \right\} \right>\\
&= \left<  \frac{1}{N} \Tr \int ds \, dt ~ {\cal P} e^{\int E }
{\cal O}_1(s) {\cal P} e^{\int E} {\cal O}_1(t)
{\cal P} e^{\int E} \right>\\
&- \left< \frac{1}{16} \tr \frac{1}{N} \Tr
\int ds \,dt ~ {\cal P} e^{\int E} \,
\xi_i(s) {\bar \psi}(x(s)) \gamma^i \, {\cal P} e^{\int E} \,
{\dot \xi}_j(t) (i\gamma^0 + \Gamma \cdot \theta) \gamma^j \psi(x(t))
\,{\cal P} e^{\int E}\right>
\end{split}\label{super}
\end{equation}

\noindent and so we have found a fermionic representation of the
first term in (\ref{2opins}). This affords a considerable
simplification of the Feynman diagrams involved in calculating the
expectation values because the fermions have less couplings than
the gauge fields and scalars.  For example, in the second
appendix, we give the computation of the leading term in the wavy
line, which is now just the free field limit of (\ref{super}).

Applying the same argument to the operator ${\cal O}_2$

\begin{equation}
\begin{split}
0 &= \frac{i}{4} \tr \left<\left\{Q_p,
\frac{1}{N} \Tr \int ds\, {\cal P} e^{ \int_s^\infty ds_1
E(s_1) } \,\xi_i(s) \, \xi_j(s) [ {\bar Q}, D_{(i}A_{j)} ](s) \, {\cal P} e^{ \int_{ - \infty }^s ds_2 E(s_2) }
\right\}\right>\\
&= \left<
\frac{1}{N} \Tr \int ds\, {\cal P} e^{ \int_s^\infty ds_1
E(s_1) } \,{\cal O}_2(s)\, {\cal P} e^{ \int_{ - \infty }^s ds_2 E(s_2) }
\right>
\end{split}
\end{equation}

\noindent we find that ${\cal O}_2$ does not contribute at any order in
perturbation theory. As is explained in \cite{Polyakov:2000ti}, the loop
operator $\hat L$ acting on the Wilson loop $W$ is defined as the coefficient
of $\delta(s-s')$ in the expression for

\begin{equation}
\frac{\delta^2 W}{\delta x_\mu(s) \delta x_\mu (s')}
\end{equation}

\noindent According to (\ref{2opins}), we have

\begin{equation}
{\hat L}\, W = \left<\frac{1}{N} \Tr {\cal P} e^{ \int E} \left[\,
i D_j F_{j 0}(x(s)) + D_j D_j \Phi(x(s)) \cdot \theta \,\right]
{\cal P} e^{ \int E}\right> ~~=~~0
\end{equation}

The infinite straight line is a solution of the loop equation.
Note that in this case it is not a simple consequence of the
equations of motion, as the potential terms for scalars and the
fermionic currents are absent.  It is, on the other hand, a result
of the 1/2 BPS nature of the Wilson loop. It would be interesting
to see whether other partially supersymmetric loops
\cite{Zarembo:2002an} also obey the loop equation.

\section*{Acknowledgments}
This work is supported in part by NSERC of Canada.  The work of
D.Y. is supported by an NSERC Postgraduate Scholarship.  Our
interests in wavy lines was inspired by conversations with V.
Kazakov, A. Polyakov and recently with Soo-Jong Rey.

\appendix
\section{Conventions}
The action of ${\cal N}=4$ supersymmetric Yang-Mills theory that
we use for our perturbative computations is
$$
S=\int d^{2\omega}x \frac{1}{2g^2} \left\{\frac{1}{2}
(F^a_{\mu\nu})^2+ (\partial_\mu\Phi^a+f^{abc}A^b_\mu\Phi^c)^2
+\bar\psi^ai\gamma^\mu
\partial_\mu\psi^a+
\right.
$$
$$
\left.~~~~~~~~~~~~~~~~~~~~~~~~~~~~~~~~~~~~~~~~~~~~~~~~~
+\sum_{I<J=1}^{10-2\omega}
\left(f^{abc}\Phi^{bI}\Phi^{Jc}\right)^2+f^{abc}\bar\psi^ai\left(\gamma^\mu
A_\mu^b+\Gamma^I \Phi^{bI}\right)\psi^c\right\}
$$
where
$$
F^a_{\mu\nu}=\partial_\mu A^a_\nu-\partial_\nu
A^a_\mu+f^{abc}A^b_\mu A^c_\nu
$$
All variables are $N\times N$ Hermitian matrices which can be
expanded in SU(N) Lie algebra generators,
$$A_\mu(x)=\sum_{a=1}^{N^2-1}A_\mu^a T^a $$ The
generators are normalized so that
$$
{\rm Tr}\left( T^a T^b\right)=\frac{1}{2}\delta^{ab}
$$
We use regularization by dimensional reduction where use
10-dimensional ${\cal N}=1$ supersymmetric Yang-Mills theory
dimensionally reduced to $2\omega$ spacetime dimensions. There are
$2\omega$ components of the vector field $A_\mu$ and $10-2\omega$
scalar fields $\Phi^I$. The fermions always have 16 real
components and the Dirac matrices are appropriate to a
Majorana-Weyl spinor in 10-dimensions.

 All of our computations are
done in the Feynman gauge where the free field correlation
functions are
$$
\left<A^a_{\mu}(x) A^b_{\nu }(y)\right>_0=g^2\Delta(x-y)
\delta_{\mu\nu}\delta^{ab} ~~,~~\left<
\Phi^{Ia}(x)\Phi^{Jb}(y)\right>_0=g^2\Delta(x-y)
\delta^{IJ}\delta^{ab} ~~, $$ $$ \left<
\psi^a(x)\bar\psi^b(y)\right>_0=g^2
i\gamma^\mu\partial_\mu\Delta(x-y)\delta^{ab}
$$
where
$$
\Delta(x)=\frac{ \Gamma(\omega-1)}{4\pi^\omega}\frac{1}{
[(x-y)^2]^{\omega-1} }
$$
We use dimensional reduction of ${\cal N}=1$ supersymmetric
Yang-Mills theory in 10-dimensions to $2\omega$-dimensions.  The
physical dimension is $2\omega=4$.
 Note the factors of the coupling constant, which come from our
normalization of the action.

\section{Example}

Consider the wavy line written in the form
\begin{eqnarray}
&&\left<W(-)\right>= 1+  \nonumber \\
&+&\frac{1}{32N}\int dsdt\left< \Tr {\cal P} e^{\int
E}\bar\psi(x(s)) \gamma\cdot\xi(s)  e^{\int E} (i\gamma^0 + \Gamma
\cdot \theta) \gamma\cdot\dot\xi(t) \psi(x(t)) e^{\int
E}\right> \nonumber \\
&+&\ldots
\end{eqnarray}
where $x(s)=(s,0,0,0)$.  In the leading order, we insert the free
field fermion propagator:
\begin{equation}
\left<W(-)\right>=1+\frac{g^2N}{256\pi^2}\int dsdt\xi_i(s)\dot
\xi_j(t) \Tr \left[\gamma^i  \gamma^j (i\gamma^0 + \Gamma \cdot
\theta)(-i\gamma^0)\right]\frac{d}{ds}\frac{1}{(s-t)^2}+\ldots
\end{equation}
which can be written as
\begin{eqnarray}
\left<W(-)\right>=1+\frac{g^2N}{16\pi^2}\int dsdt\xi(s)\cdot\dot
\xi(t)\frac{d}{ds}
 \frac{1}{(t-s)^2}+\ldots  \nonumber \\
 =1+\frac{g^2N}{16\pi^2}\int dsdt\frac{[\dot\xi(s)-\dot\xi(t)]^2}{2(s-t)^2} +\ldots
\end{eqnarray}
\end{fmffile}
which is identical to our previous result.


\begin{thebibliography}{99}

\bibitem{Maldacena:1997re}
J.~M.~Maldacena, ``The large N limit of superconformal field
theories and supergravity,'' Adv.\ Theor.\ Math.\ Phys.\  {\bf 2},
231 (1998) [Int.\ J.\ Theor.\ Phys.\  {\bf 38}, 1113 (1999)]
[arXiv:hep-th/9711200].



\bibitem{Gubser:1998bc}
S.~S.~Gubser, I.~R.~Klebanov and A.~M.~Polyakov, ``Gauge theory
correlators from non-critical string theory,'' Phys.\ Lett.\ B
{\bf 428}, 105 (1998) [arXiv:hep-th/9802109].


\bibitem{Witten:1998qj}
E.~Witten, ``Anti-de Sitter space and holography,'' Adv.\ Theor.\
Math.\ Phys.\  {\bf 2}, 253 (1998) [arXiv:hep-th/9802150].

\bibitem{Polyakov:1998ju}
A.~M.~Polyakov, ``The wall of the cave,'' Int.\ J.\ Mod.\ Phys.\ A
{\bf 14}, 645 (1999) [arXiv:hep-th/9809057].


\bibitem{Petersen:1999zh}
J.~L.~Petersen, ``Introduction to the Maldacena conjecture on
AdS/CFT,'' Int.\ J.\ Mod.\ Phys.\ A {\bf 14}, 3597 (1999)
[arXiv:hep-th/9902131].

\bibitem{DiVecchia:1999yr}
P.~Di Vecchia, ``Large N gauge theories and ADS/CFT
correspondence,'' arXiv:hep-th/9908148.

\bibitem{Akhmedov:1999rc}
E.~T.~Akhmedov, ``Introduction to the AdS/CFT correspondence,''
arXiv:hep-th/9911095.


\bibitem{Klebanov:2000me}
I.~R.~Klebanov, ``TASI lectures: Introduction to the AdS/CFT
correspondence,'' arXiv:hep-th/0009139.



\bibitem{Aharony:1999ti}
O.~Aharony, S.~S.~Gubser, J.~M.~Maldacena, H.~Ooguri and Y.~Oz,
``Large N field theories, string theory and gravity,'' Phys.\
Rept.\  {\bf 323}, 183 (2000) [arXiv:hep-th/9905111].

\bibitem{Maldacena:1998im}
J.~M.~Maldacena, ``Wilson loops in large N field theories,''
Phys.\ Rev.\ Lett.\  {\bf 80}, 4859 (1998) [arXiv:hep-th/9803002].


\bibitem{Rey:1998ik}
S.~J.~Rey and J.~T.~Yee, ``Macroscopic strings as heavy quarks in
large N gauge theory and  anti-de Sitter supergravity,'' Eur.\
Phys.\ J.\ C {\bf 22}, 379 (2001) [arXiv:hep-th/9803001].

\bibitem{Rey:1998bq}
S.~J.~Rey, S.~Theisen and J.~T.~Yee, ``Wilson-Polyakov loop at
finite temperature in large N gauge theory and anti-de Sitter
supergravity,'' Nucl.\ Phys.\ B {\bf 527}, 171 (1998)
[arXiv:hep-th/9803135].

\bibitem{Danielsson:1998br}
U.~H.~Danielsson and A.~P.~Polychronakos, ``Quarks, monopoles and
dyons at large N,'' Phys.\ Lett.\ B {\bf 434}, 294 (1998)
[arXiv:hep-th/9804141].


\bibitem{Gross:1998gk}
D.~J.~Gross and H.~Ooguri, ``Aspects of large N gauge theory
dynamics as seen by string theory,'' Phys.\ Rev.\ D {\bf 58},
106002 (1998) [arXiv:hep-th/9805129].

\bibitem{Drukker:1999zq}
N.~Drukker, D.~J.~Gross and H.~Ooguri, ``Wilson loops and minimal
surfaces,'' Phys.\ Rev.\ D {\bf 60}, 125006 (1999)
[arXiv:hep-th/9904191].


\bibitem{Greensite:1998bp}
J.~Greensite and P.~Olesen, ``Remarks on the heavy quark potential
in the supergravity approach,'' JHEP {\bf 9808}, 009 (1998)
[arXiv:hep-th/9806235].

\bibitem{Witten:jd}
E.~Witten, ``Branes And The Dynamics Of QCD,'' Nucl.\ Phys.\
Proc.\ Suppl.\  {\bf 68}, 216 (1998).

\bibitem{Aharony:1998qu}
O.~Aharony and E.~Witten, ``Anti-de Sitter space and the center of
the gauge group,'' JHEP {\bf 9811}, 018 (1998)
[arXiv:hep-th/9807205].



\bibitem{Kinar:1998vq}
Y.~Kinar, E.~Schreiber and J.~Sonnenschein, ``Q anti-Q potential
from strings in curved spacetime: Classical results,'' Nucl.\
Phys.\ B {\bf 566}, 103 (2000) [arXiv:hep-th/9811192].


\bibitem{Danielsson:1998wt}
U.~H.~Danielsson, E.~Keski-Vakkuri and M.~Kruczenski, ``Vacua,
propagators, and holographic probes in AdS/CFT,'' JHEP {\bf 9901}
(1999) 002 [arXiv:hep-th/9812007].

\bibitem{Zarembo:2001jp}
K.~Zarembo, ``String breaking from ladder diagrams in
super-Yang-Mills theory theory,'' JHEP {\bf 0103}, 042 (2001)
[arXiv:hep-th/0103058].

\bibitem{Polchinski:2000uf}
J.~Polchinski and M.~J.~Strassler, ``The string dual of a
confining four-dimensional gauge theory,'' arXiv:hep-th/0003136.



\bibitem{Janik:2000pp}
R.~A.~Janik, ``String fluctuations, AdS/CFT and the soft pomeron
intercept,'' Phys.\ Lett.\ B {\bf 500}, 118 (2001)
[arXiv:hep-th/0010069].


\bibitem{Polchinski:2001tt}
J.~Polchinski and M.~J.~Strassler, ``Hard scattering and
gauge/string duality,'' Phys.\ Rev.\ Lett.\  {\bf 88}, 031601
(2002) [arXiv:hep-th/0109174].


\bibitem{Janik:2001sc}
R.~A.~Janik and R.~Peschanski, ``Reggeon exchange from AdS/CFT,''
Nucl.\ Phys.\ B {\bf 625}, 279 (2002) [arXiv:hep-th/0110024].



\bibitem{Semenoff:2002kk}
G.~W.~Semenoff and K.~Zarembo, ``Wilson loops in super-Yang-Mills
theory theory: From weak to strong coupling,'' Nucl.\ Phys.\
Proc.\ Suppl.\  {\bf 108}, 106 (2002) [arXiv:hep-th/0202156].

\bibitem{Zarembo:2002an}
K.~Zarembo, ``Supersymmetric Wilson loops,'' Nucl.\ Phys.\ B {\bf
643}, 157 (2002) [arXiv:hep-th/0205160].


\bibitem{Polchinski:2002jw}
J.~Polchinski and M.~J.~Strassler, ``Deep inelastic scattering and
gauge/string duality,'' JHEP {\bf 0305}, 012 (2003)
[arXiv:hep-th/0209211].

\bibitem{Kruczenski:2002fb}
M.~Kruczenski, ``A note on twist two operators in N = 4 SYM and
Wilson loops in Minkowski signature,'' JHEP {\bf 0212}, 024 (2002)
[arXiv:hep-th/0210115].


\bibitem{Makeenko:2002qe}
Y.~Makeenko, ``Light-cone Wilson loops and the string / gauge
correspondence,'' JHEP {\bf 0301}, 007 (2003)
[arXiv:hep-th/0210256].





\bibitem{Mikhailov:2002ya}
A.~Mikhailov, ``Special contact Wilson loops,''
arXiv:hep-th/0211229.

\bibitem{Kotikov:2003fb}
A.~V.~Kotikov, L.~N.~Lipatov and V.~N.~Velizhanin, ``Anomalous
dimensions of Wilson operators in N = 4 super-Yang-Mills theory
theory,'' Phys.\ Lett.\ B {\bf 557}, 114 (2003)
[arXiv:hep-ph/0301021].


\bibitem{Belitsky:2003ys}
A.~V.~Belitsky, A.~S.~Gorsky and G.~P.~Korchemsky, ``Gauge /
string duality for QCD conformal operators,'' Nucl.\ Phys.\ B {\bf
667}, 3 (2003) [arXiv:hep-th/0304028].

\bibitem{Mikhailov:2003er}
A.~Mikhailov, ``Nonlinear waves in AdS/CFT correspondence,''
arXiv:hep-th/0305196.

\bibitem{Arutyunov:2003uj}
G.~Arutyunov, S.~Frolov, J.~Russo and A.~A.~Tseytlin, ``Spinning
strings in AdS(5) x S**5 and integrable systems,'' Nucl.\ Phys.\ B
{\bf 671}, 3 (2003) [arXiv:hep-th/0307191].


\bibitem{Gorsky:2003nq}
A.~Gorsky, ``Spin chains and gauge / string duality,''
arXiv:hep-th/0308182.

\bibitem{Belitsky:2003sh}
A.~V.~Belitsky, S.~E.~Derkachov, G.~P.~Korchemsky and
A.~N.~Manashov, ``Superconformal operators in N = 4
super-Yang-Mills theory,'' arXiv:hep-th/0311104.

\bibitem{Kruczenski:2003wz}
M.~Kruczenski, ``Wilson loops and anomalous dimensions in
cascading theories,'' Phys.\ Rev.\ D {\bf 69}, 106002 (2004)
[arXiv:hep-th/0310030].

\bibitem{Pestun:2002mr}
V.~Pestun and K.~Zarembo, ``Comparing strings in AdS(5) x S**5 to
planar diagrams: An example,'' Phys.\ Rev.\ D {\bf 67}, 086007
(2003) [arXiv:hep-th/0212296].



\bibitem{Bianchi:2002gz}
M.~Bianchi, M.~B.~Green and S.~Kovacs, ``Instanton corrections to
circular Wilson loops in N = 4 supersymmetric Yang-Mills,'' JHEP
{\bf 0204}, 040 (2002) [arXiv:hep-th/0202003].


\bibitem{Kotikov:2004er}
A.~V.~Kotikov, L.~N.~Lipatov, A.~I.~Onishchenko and
V.~N.~Velizhanin, ``Three-loop universal anomalous dimension of
the Wilson operators in N = 4
arXiv:hep-th/0404092.



\bibitem{Erickson:2000af}
J.~K.~Erickson, G.~W.~Semenoff and K.~Zarembo, ``Wilson loops in N
= 4 supersymmetric Yang-Mills theory,'' Nucl.\ Phys.\ B {\bf 582},
155 (2000) [arXiv:hep-th/0003055].


\bibitem{Plefka:2001bu}
J.~Plefka and M.~Staudacher, ``Two loops to two loops in N = 4
supersymmetric Yang-Mills theory,'' JHEP {\bf 0109}, 031 (2001)
[arXiv:hep-th/0108182].

\bibitem{Arutyunov:2001hs}
G.~Arutyunov, J.~Plefka and M.~Staudacher, ``Limiting geometries
of two circular Maldacena-Wilson loop operators,'' JHEP {\bf
0112}, 014 (2001) [arXiv:hep-th/0111290].


\bibitem{Drukker:2000rr}
N.~Drukker and D.~J.~Gross, ``An exact prediction of N = 4
supersymmetryM theory for string theory,'' J.\ Math.\ Phys.\  {\bf
42}, 2896 (2001) [arXiv:hep-th/0010274].

\bibitem{Semenoff:2001xp}
G.~W.~Semenoff and K.~Zarembo, ``More exact predictions of SUSYM
for string theory,'' Nucl.\ Phys.\ B {\bf 616}, 34 (2001)
[arXiv:hep-th/0106015].

\bibitem{Polyakov:2000ti}
A.~M.~Polyakov and V.~S.~Rychkov,
``Gauge fields - strings duality and the loop equation,''
Nucl.\ Phys.\ B {\bf 581}, 116 (2000)
[arXiv:hep-th/0002106].

\bibitem{Kazakov}V.~Kazakov, private communication.

\bibitem{Polyakov:2000jg}
A.~M.~Polyakov and V.~S.~Rychkov, ``Loop dynamics and AdS/CFT
correspondence,'' Nucl.\ Phys.\ B {\bf 594}, 272 (2001)
[arXiv:hep-th/0005173].

\bibitem{Makeenko:pb}
Y.~M.~Makeenko and A.~A.~Migdal, ``Exact Equation For The Loop
Average In Multicolor QCD,'' Phys.\ Lett.\ B {\bf 88}, 135 (1979)
[Erratum-ibid.\ B {\bf 89}, 437 (1980)].

\bibitem{Makeenko:vm}
Y.~Makeenko and A.~A.~Migdal, ``Quantum Chromodynamics As Dynamics
Of Loops,'' Nucl.\ Phys.\ B {\bf 188}, 269 (1981) [Sov.\ J.\
Nucl.\ Phys.\ {\bf 32}, 431.1980\ YAFIA,32,838 (1980\
YAFIA,32,838-854.1980)].

\bibitem{Berenstein:1998ij}
D.~Berenstein, R.~Corrado, W.~Fischler and J.~M.~Maldacena, ``The
operator product expansion for Wilson loops and surfaces in the
large N
Phys.\ Rev.\ D {\bf 59}, 105023 (1999) [arXiv:hep-th/9809188].




\end{thebibliography}
\end{document}